\documentclass[%
 reprint,superscriptaddress,
 amsmath,amssymb,
 aps,
]{revtex4-2}

\newcommand{\mSR}{$\mu$SR\xspace}
\newcommand{\DFTmu}{DFT+$\mu$\xspace}
\newcommand{\Si}{V$_{3}$Si\xspace}
\newcommand{\Nb}{Nb$_{3}$Sn\xspace}
\newcommand{\Sn}{V$_{3}$Sn\xspace}

\usepackage{siunitx}
\usepackage{nicefrac}
\usepackage{chngpage}
\usepackage{amssymb}
\usepackage{amsmath}
\usepackage{appendix}
\usepackage{latexsym}
\usepackage{listings}
\usepackage{amsthm}
\usepackage{eucal}
\usepackage{xspace}

\usepackage{eufrak}
\usepackage{graphicx}
\usepackage[%
    colorlinks=true,
    pdfborder={0 0 0},
    linkcolor=red
]{hyperref}
\usepackage{dcolumn}
\usepackage{bm}

\begin{document}
\graphicspath{ {figures/} }
\preprint{UPROXUBO-03}

\title{Entanglement between a muon spin and $I>\frac{1}{2}$ nuclear spins}

\author{Pietro Bonf\`a}%
 \email{pietro.bonfa@unipr.it
 }
\affiliation{%
Department of Mathematical, Physical and Computer Sciences, University of Parma, Italy
}

\author{Jonathan Frassineti}
 \affiliation{Department of Physics and Astronomy, University of Bologna, Italy}

\author{John M. Wilkinson}
\affiliation{
Clarendon Laboratory, Department of Physics, University of Oxford, Parks Road, Oxford OX1 3PU, United Kingdom
}

\author{Giacomo Prando}
\affiliation{%
Department of Physics, University of Pavia, Italy
}%

\author{Muhammad Maikudi Isah}%
\affiliation{%
Department of Mathematical, Physical and Computer Sciences, University of Parma, Italy
}%

\author{Chennan Wang}
\affiliation{%
Laboratory for Muon Spin Spectroscopy, Paul Scherrer Institute, CH-5232 Villigen, Switzerland
}%
 
\author{Tiziana Spina}
\affiliation{%
Superconducting Radio Frequency (SRF) Materials and Research Department, Fermilab, Batavia, USA
}

\author{Boby Joseph}
\affiliation{%
Elettra-Sincrotrone Trieste, S.S. 14-km 163.5, Basovizza, 34149 Trieste, Italy
}%

\author{Vesna F. Mitrovi\'c}
\affiliation{
Department of Physics, Brown University, Providence, 02912 Rhode Island, USA
}%

\author{Roberto De Renzi}
\affiliation{%
Department of Mathematical, Physical and Computer Sciences, University of Parma, Italy\\
}%

\author{Stephen J. Blundell}
\affiliation{
Clarendon Laboratory, Department of Physics, University of Oxford, Parks Road, Oxford OX1 3PU, United Kingdom 
}%

\author{Samuele Sanna}
\affiliation{Department of Physics and Astronomy, University of Bologna, Italy}

\date{\today}

\begin{abstract}
We report on the first example of quantum coherence between the spins of
muons and quadrupolar nuclei. We observe this effect in vanadium intermetallic
compounds which adopt the A15 crystal structure, and whose members include all
technologically dominant superconductors. The entangled states are extremely
sensitive to the local structural and electronic environments through the
electric field gradient at the quadrupolar nuclei. This case-study demonstrates 
that positive muons can be used as a quantum sensing tool to probe also
structural and charge related phenomena in materials, even in the absence 
of magnetic order.
\end{abstract}

\maketitle

Quantum coherence between an implanted positively-charged muon and
nuclei in a solid was first conclusively demonstrated using muon-spin
spectroscopy (\mSR) experiments on simple ionic fluorides
\cite{Brewer86}.  The strong hydrogen-like bonding of the implanted
positive muon (chemically identified as $\mu^{+}$) to nearest-neighbor F
ions, characterized by a single spin $1/2$ $^{19}$F nuclear
isotope, gives rise to a hierarchical separation of the muon spin
interactions. Typically, dipolar couplings with two nearest-neighbor
(nn) $^{19}$F nuclear spins, $\boldsymbol I_1$ and $\boldsymbol I_2$,
determine the dominant spin-Hamiltonian of the $S=1/2$ muon,
whereas all the residual interactions, starting from the next nearest
neighbors (nnn), can be ignored to a first approximation. Thanks to
the 100\% initial muon spin polarization, a prerogative of \mSR, this
shows up experimentally as a characteristic coherent spin precession
pattern in the muon time-dependent asymmetry, uniquely determined by
the geometry of the F--$\mu$--F bonds. Many fluorinated
compounds display this coherent pattern in non-magnetic phases,
including ionic fluorides
\cite{Noakes1993,Lancaster2007b,Lancaster2009}, fluropolymers
\cite{Pratt2003,Nishiyama2003} and molecular magnets
\cite{Lancaster2007a}.  For these materials, the absence or the fast
fluctuation of electronic magnetic moments leave the nuclear spin
interactions to determine the dynamics of the muon spin
polarization. This allows a very precise assignment of the muon
implantation site, now known to be particularly accurate with the help
of density functional theory (DFT) {\it ab-initio} simulations of the
muon stopping-site inside the crystal (a technique which is also known
as \DFTmu\ \cite{Bernardini13, Moeller12, Moeller13, Bonfa16,blundell_renzi_lancaster_pratt_2022}). A
similar coherent spin behavior has been identified in certain hydrides
\cite{Lord2000,Mendels2007,Kadono08} and in metal–organic frameworks \cite{acs.nanolett.0c03140}, where for instance a close
association of a proton and the positive muon approximates a muoniated
hydrogen molecule, $\mu$H, or possibly, a bonded molecular ion,
($\mu$H)$^+$, ($\mu$H)$^-$.
Notice that $^1$H, like $^{19}$F, is a spin $I=1/2$ nucleus.

In the case of H, as for the cases of many other nuclear species, such
a coherent pattern is rarely observed in \mSR  experiments. Much
more often a large number of unpolarized nuclear spins give rise to a
$T_2^{-1}$ relaxation process with either Gaussian or Lorentzian
lineshapes, both the hallmarks of fast decoherence on the timescale of
the period of the coherent quantum interference processes.  Fluorine
is special since it is very electronegative, and it has both a small
ionic radius and a large nuclear moment, so that its dipolar coupling
to the muon is strong and consequently several oscillations in any
quantum-coherent signal can be observed before all muons have decayed
or any nuclear relaxation process has become significant.  The special
F--$\mu$--F case was very recently revisited by some of us \cite
{PhysRevLett.125.087201}, showing the role of the rest of the nuclear spins (nnn
and beyond) in the slow decoherence process of F--$\mu$--F.  This work
implies that the very well known F--$\mu$--F effect, confined until now
among the technicalities of the muon spectroscopy, displays all the
features of a very high accuracy {\it quantum sensor} that can be
exploited for microscopic detection of important physical
phenomena \cite{wilkinson2021muon}. Unfortunately, until now, the sensor
has been available only for
F$^-$- and, much more seldom, for H$^-$-containing materials.

 \begin{figure*}[th]
     \centering
     \includegraphics{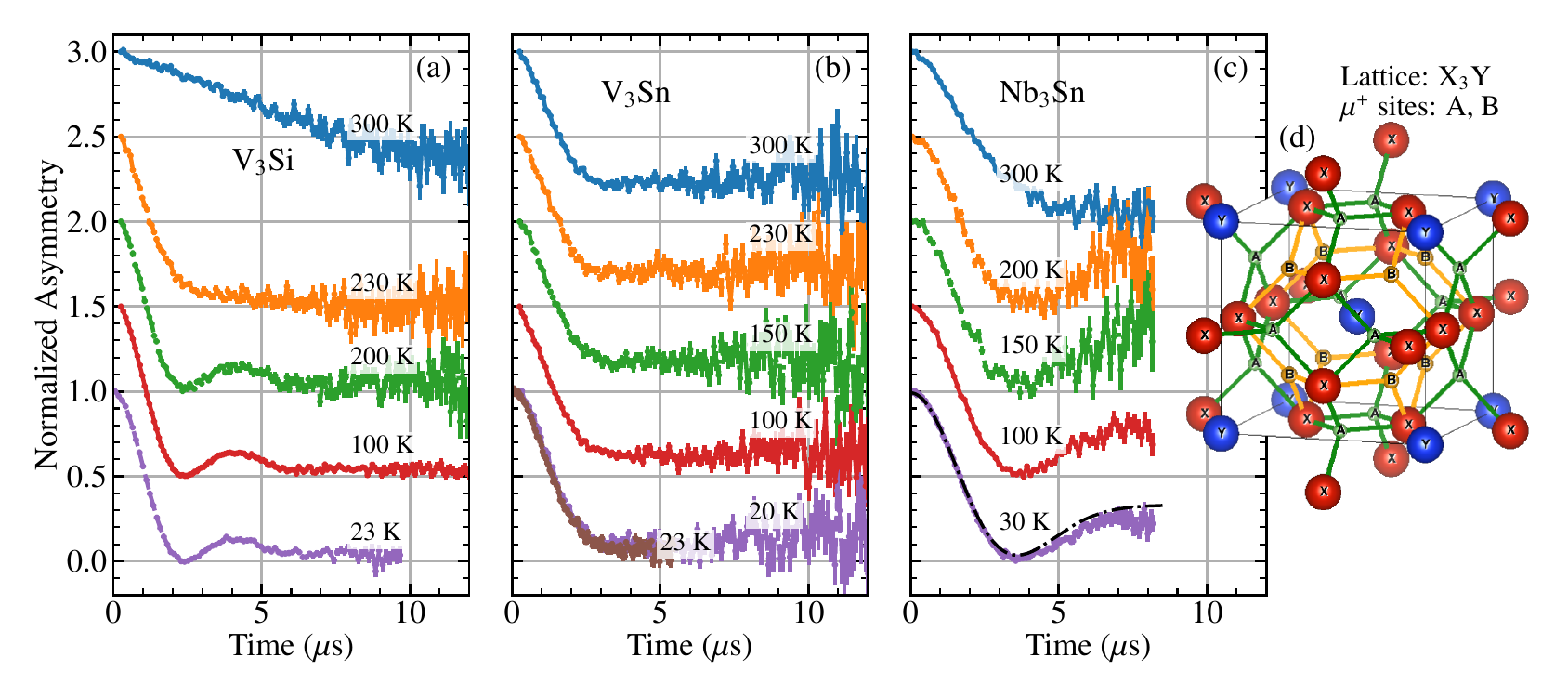}
     \caption{Experimental results obtained for \Si (a), \Sn (b) and \Nb (c)
     at various temperatures in ZF. The initial asymmetry has been
     normalized to 1 and and the various measurements are shifted along
     the $y$ axis by multiples of 0.5. 
     The black line in (c) is a best fit to a static Kubo-Toyabe function.
     In picture (d) the lattice structure of A15 compounds and the candidate muon sites
     identified in this class of materials are depicted.}
     \label{fig:experimental}
\end{figure*}

In the present work we demonstrate the same surprising type of quantum
coherence due to the entanglement of the muon spin with nn 
nuclear spin in the case of $I>1/2$. We show this phenomenon in three
intermetallic compounds,
Nb$_3$Sn, V$_3$Si 
and V$_3$Sn, which belong to the A15 cubic phases (Pm$\overline 
3$n, group number 223), whose members include several technologically 
dominant conventional superconductors \cite{Muller80}. 
In stark contrast to the well-studied $I=1/2$ case of $^{19}$F 
and $^1$H, the presence of nn nuclei with $I>1/2$, namely $I= 7/2$, 
$9 / 2$ of $^{51}V$ and $^{93}$Nb respectively, implies the existence of 
quadrupolar interactions. This has two effects that could potentially
spoil the {\it quantum sensor} concept: first, it was until now
unclear that a detectable quantum coherence could nevertheless show up
in the muon asymmetry; second, quadrupolar interactions are
proportional to the electric field gradient (EFG) at the nucleus in
question, not just on the pure geometry of the bonds. EFG tensors
are very accurately determined by DFT in bulk materials \cite{Blaha1985} 
and compared to the values measured for instance by nuclear
magnetic resonance (NMR). The muon embedding in the crystal
alters the bulk EFG in more than one way. We show that the coherent
effect survives and we develop here an accurate model to describe this 
phenomenon. Our modeling of the coherence entails identifying precisely 
the muon site and calculating muon perturbed EFG tensors at nn and nnn 
nuclei. The results show that the observed phenomenon is highly sensitive 
to small structural and electronic differences among the same A15 family,
paving the way to extend the use of muon spectroscopy as a quantum 
sensing technique for charge-related phenomena.

Zero-field (ZF) $\mu$SR temperature scans, using the EMU spectrometer at the ISIS Muon Source and
the GPS spectrometer \cite{GPS} at the Paul Scherrer Institute, have been conducted as a function of temperature.
Further details on the experimental methods are provided in the
Supplemental Material (SM)\footnote{See Supplemental Material at 
[URL will be inserted by publisher] for additional details concerning 
the experimental setting and the results of DFT simulations. 
Input and output data to reproduce our results are available via Materials Cloud \cite{archive}.}.
Fig.~\ref{fig:experimental} shows the $\mu$SR spectra (time-dependent spin polarization of the muon ensemble) 
for all the samples at various representative temperatures. 
The temperature dependence is relatively weak, except above $200$~K, where 
thermally activated $\mu^{+}$ diffusion occurs
\cite{Yaouanc1986}.
At low temperature, where the muon is static in the \mSR time window,
the results are remarkably sample dependent despite all the X$_3$Y samples
($X=\{\mathrm{V,Nb}\}$ and $Y=\{\mathrm{Si,Sn}\}$) being very similar metals,
sharing the same A15 cubic lattice structure.
The structure is shown in Fig.~\ref{fig:experimental}d and our samples 
have a cubic lattice parameter $a=$\SI{4.72}{\angstrom}, \SI{4.98}{\angstrom},
\SI{5.29}{\angstrom} for \Si, \Sn and \Nb respectively (see  \cite{Note1}),
in agreement with previous results \cite{Paduani_2008, Morton_1979,Alimenti_2020}. 
The nuclei of the X atoms are closer to the calculated muon sites,
as shown in Fig.~\ref{fig:experimental}d with labels A and B, and all have 
similar properties: $^{51}$V with 99.8\% abundance has spin $I=7/2$, 
gyromagnetic ratio $\gamma_{V}=70.45\times10^{6}$~rad/(sT) and 
quadrupole moment $Q=$-0.052(10) barn and $^{93}$Nb with 100\% abundance 
has spin $I=9/2$, $\gamma_{Nb}=65.64\times10^{6}$~rad/(sT) and $Q=$-0.32(2) barn 
\footnote{Multiple inconsistent results have been published for the quadrupole moments of $^{51}$V and $^{94}$Nb.
We used recommended values from IAEA available at
\url{https://www-nds.iaea.org/nuclearmoments/isotope_measurement_results.php?A=51&Z=23}
and \url{ https://www-nds.iaea.org/nuclearmoments/isotope_measurement_results.php?A=93&Z=41}}.

\begin{figure*}
     \centering
     \includegraphics{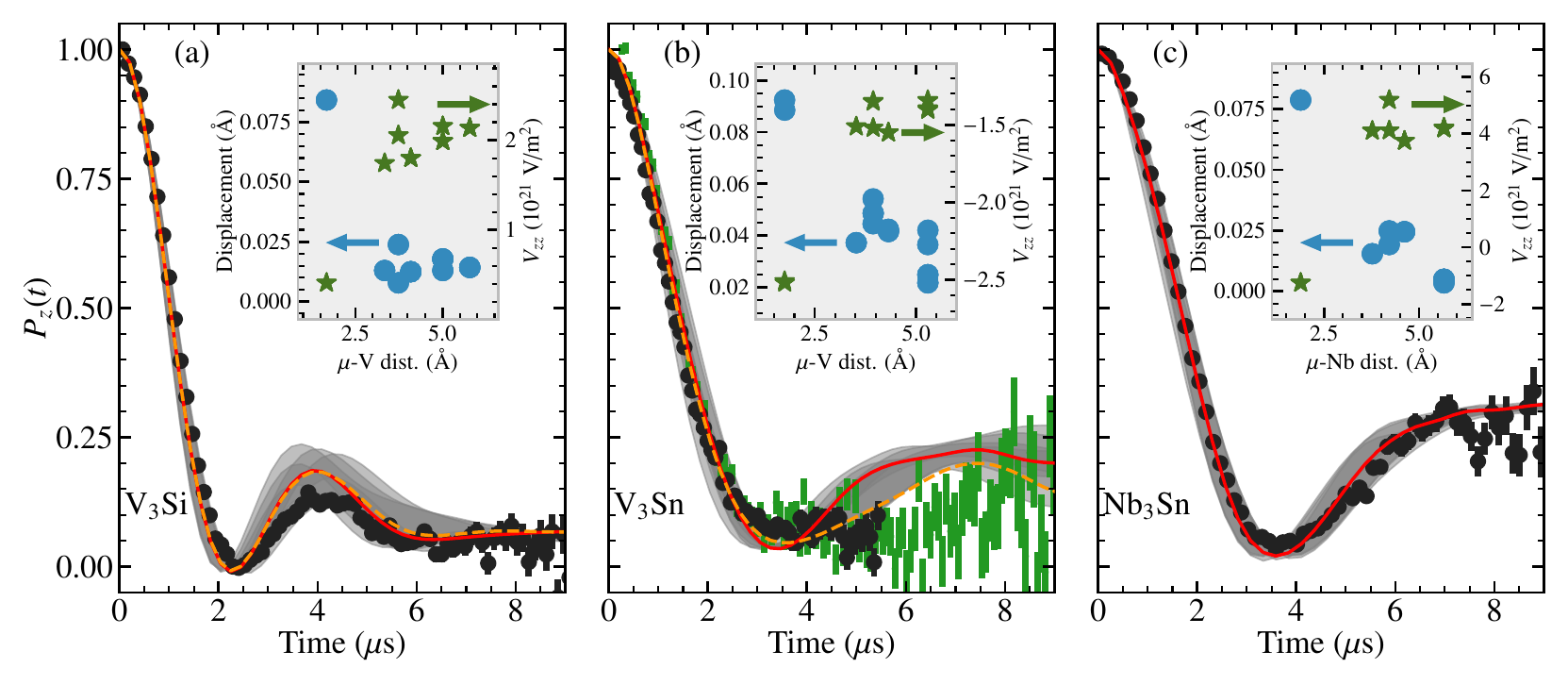}
     \caption{Comparison between experimental and predicted muon spin 
     polarization obtained using atomic displacements and EFGs 
     from plane wave based DFT calculations. The black dots in panels a), b) and c) 
     are lowest temperature data collected at PSI for \Si, \Sn and \Nb respectively. 
     The green bars in panel (b) are ISIS results collected at 20~K. 
     A background has been estimated by comparing the asymmetries collected at ISIS and PSI and removed. 
     The red (orange) line in all plots is the depolarization obtained 
     using first principles results from PW (FP) simulations to solve Eq.~\ref{eq:model}. 
     Shaded area highlight different trends that originate by taking 
     into account typical uncertainties of the DFT based predictions (see main text).
     The insets show the perturbation induced by the muon on its X-type neighbours ($X=$V,Nb). 
     In particular, in the presence of the muon, the displacement of 
     each $X$ atom from its equilibrium position in the unperturbed 
     lattice and the values of $V_{zz}$ at the considered atomic site 
     are reported on the left-hand and right-hand y-axes, respectively, 
     as a function of the unperturbed distance of the considered atom 
     from the $\mu^+$ interstitial position in a 3x3x3 supercell.
     \label{fig:model}}
\end{figure*}

The oscillatory behavior observed in \Si (Fig.~\ref{fig:experimental}a) is in marked 
contrast to the cases of both \Nb (Fig.~\ref{fig:experimental}c), which resembles 
the conventional Kubo-Toyabe (KT) relaxation function (empirical KT best 
fit shown by the dashed line in the same panel and characterized by a 
dip and a tail that flattens at $\nicefrac{1}{3}$ of the initial value), and of \Sn  
(Fig.~\ref{fig:experimental}b), which could be described by a KT relaxation, 
with an additional decay of the $\nicefrac{1}{3}$ tail which has no evident physical origin.
The surprisingly slow oscillations observed in \Si (Fig.~\ref{fig:experimental}a) 
cannot be due to internal fields of electronic origin since all these A15 samples are non-magnetic. 
Instead, as we will show, they result from a quantum coherent precession 
pattern due to the coupling between the muon and nearby $^{51}$V nuclear moments, 
analogous to the F--$\mu$--F case, and never reported before for systems containing $I > 1/2$ nuclear spins.

In order to explain the three precession patterns of Fig.~\ref{fig:experimental}  
we now consider the microscopic nuclear and electronic degrees of 
freedom entering the quantum mechanical model of the muon polarization.
The model requires the knowledge of three ingredients to reproduce the experimental muon polarization:
(i) the muon site, 
(ii) the perturbation induced by the $\mu^{+}$ on the position of the neighboring atoms, 
(iii) the perturbation induced by the muon on the EFG at the nuclear sites with spin $I>1/2$.
These information allow to fully define the spin Hamiltonian $\mathcal{H}$ given by
\begin{equation}
      \mathcal{H} = \sum_i ^{N_{nuc}}  \frac{\mu_0}{4 \pi } \frac{\gamma_\mu \gamma_{i} \hbar^2}{r^3_{i}} \mathbf{S}_\mu \cdot \mathbf{D}^{i} \cdot  \mathbf{I}^{i} +
     \frac{eQ_{i}}{2 I(2 I-1)} \mathbf{I}^{i} \cdot \mathbf{V}^{i} \cdot \mathbf{I}^{i}, 
     \label{eq:model}
\end{equation}
where $\mathbf{S}_\mu$ is the spin of the muon and $r_{i}$ is its distance 
from nucleus $i$, $\mathbf{I}^{i}$ and $Q_{i}$ are respectively the spin 
and the quadrupole moment of nucleus $i$, $\mathbf{D}$ and $\mathbf{V}$ are 
the dipolar and EFG tensors at nuclear site $i$, and other symbols have their standard meaning.
All the quantities entering Eq.~\ref{eq:model} can be accurately estimated
with DFT-based \emph{ab initio} approaches and we describe below the 
results that we obtained following the \DFTmu procedure.

Two candidate muon sites are present in our A15 compounds and are shown in Fig.~\ref{fig:experimental}d with labels A and B.
Site A is located in the center of the tetrahedron formed by four X atoms
while site B is in the center of the triangle formed by three X atoms.
We find that site B always has higher energy than site A by hundreds of meV (see \cite{Note1} for details)
and is therefore omitted from the subsequent analysis.
DFT simulations produce, as an additional outcome, the displacements of the
atoms surrounding the muon.
In all cases, the nn X atoms are substantially
displaced by the muon and the nearest neighbor distances increase by
$6$\%, $5$\%, $4$\% respectively in \Si, \Sn, \Nb (the absolute values are
shown in the insets of Fig.~\ref{fig:model} against the unperturbed $\mu$-X distance and in SM \cite{Note1}).

The next step is the evaluation of the EFG at the quadrupolar nuclei in each compound.
While for ionic materials a point charge approximation may sometimes be sufficient,
covalent and metallic systems require more elaborate strategies.
Full potential (FP) DFT simulations yield very accurate
estimates in materials where the mean field approximation does not break down
owing to strong correlation, but are extremely computationally demanding.
For this reason, and aiming at providing an easily adoptable approach,
we opted for an effective compromise between accuracy and speed using a plane wave basis\cite{Giannozzi2020,QE-2009,QE-2017} combined with PAW \cite{PhysRevB.50.17953} pseudopotentials.
A detailed discussion of our strategy and additional comparisons with FP simulations\cite{elk} are provided in the SM \cite{Note1}.
Notably, this procedure converges much faster than the equivalent technique aimed at the prediction of magnetic contact hyperfine fields at the muon sites \cite{PhysRevB.97.174414}.

Unsurprisingly, the EFG of the four X neighbors of the muon is drastically affected by the presence of the interstitial charge.
For example, in \Si the unperturbed EFG tensor at V nuclei in the pristine material, with $V_{zz}=2.2 \times 10^{21}$ V/m$^2$ and $\eta=0$, in agreement with the experimental value of $V_{zz}=2.37 \times 10^{21}$ V/m$^2$,  reduces by almost an order of magnitude as a consequence of the presence of the positive impurity and the lattice distortion, in agreement with earlier work
\cite{Yaouanc1986}.
Note that site assignments come with some small uncertainty, and previous investigations that can be compared with experiment
\cite{PhysRevB.89.184425,PhysRevB.93.174405,PhysRevLett.125.087201, wilkinson2021muon,PhysRevB.87.121108} reveal that a discrepancy of the order of a tenth of Angstrom is to be expected.
On the other hand, plane wave based estimations of EFGs are subject to a much larger
uncertainty of the order of 30\% and $1.17 \times 10^{21}$~V/m$^2$ in relative and absolute terms \cite{Choudhary2020}.

Having collected all parameters entering Eq.~\ref{eq:model}, we proceed 
to compute the time-dependent muon polarization numerically.
For the A15 compounds the inter-nuclear dipolar interactions can be safely
neglected\footnote{The X-X atom distance is of the order of 3 \AA, about 1.75 times larger than $d_{\mu-X}$. We have also verified that inter-nuclear interactions marginally affect the depolarization signal} thus allowing the adoption of the approach proposed by 
Celio \cite{Celio1986a,Celio1986b} and implemented in the publicly available code UNDI \cite{Bonfa2021}, which makes the estimate very quick.
Our calculations consider only effect of the nearest nuclei, but
it has recently been shown by some of us \cite{PhysRevLett.125.087201}
how to effectively include the effect of farther nuclei with an appropriate re-scaling of
second nearest neighbors interaction, allowing a substantial reduction of
the otherwise exponentially diverging dimension of the Hilbert space.
Following \cite{PhysRevLett.125.087201}, we consider 4 nn and 4 nnn whose positions are homogeneously rescaled by a small amount to compensate for the remaining nuclei (see \cite{Note1} for details).

The predicted $\mu$SR signal obtained fully \emph{ab initio}, i.e. without free parameters, is shown for all samples in Fig.~\ref{fig:model} by a red line (PW results) and a orange dashed line (FP results), while shaded
area indicate the uncertainty in the PW based prediction quantified with a reduction
or increase of 3 \% (29\%) of $d_{\mu-X}$ (EFG values).
Perfect agreement is found for \Nb [Fig.~\ref{fig:model}(c)], while for \Si [Fig.~\ref{fig:model}(a)] a small deviation is observed at about 4~$\mu$s where the first bump is slightly overestimated, although
 the experimental result falls inside the shaded area. A small increase of 15 m\AA{} in the $\mu$-V distance allows to recover perfect agreement (see \cite{Note1}). Remarkably the oscillation (the time position of minima and maxima) is very well reproduced.
\Sn is the sample showing worst agreement in the long-time tail. In this case the deviation
 can be attributed to the 
limits of the PAW approximation in reconstructing the EFG at the V sites. Indeed the FP prediction, that differs from the PW based estimate by 16\%,
improves the agreement with the experimental data.
These trends demonstrate the exquisite sensitivity of \mSR to atomic distances and EFGs.

The striking difference between
the muon asymmetries collected in a set of compounds that share the same lattice
structure, the same muon site, and similar lattice distortions may appear puzzling at first sight.
To address this point, we introduce the simple and analytically
solvable case of one muon interacting with a single nucleus of spin $I$ subject to
an axial EFG \cite{Vogel1986}.
In zero external field (ZF), the interaction depends on two parameters:

\begin{figure}
     \centering
     \includegraphics{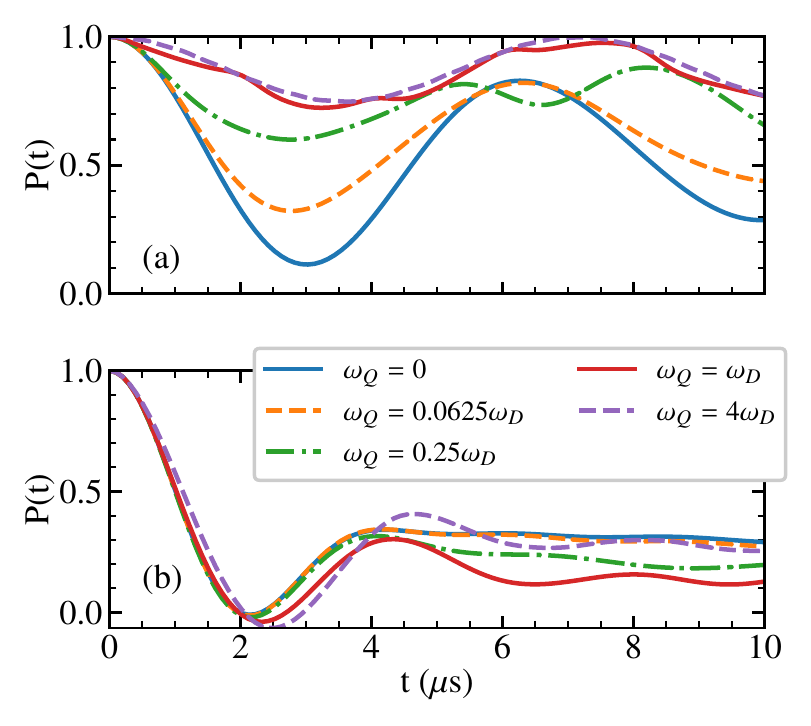}
     \caption{(a) Time-dependent spin polarization $P(t)$ for a muon interacting with a single
     nucleus with spin $I=7/2$ subject to an axial EFG for various values of $\omega_{Q}/\omega_{D}$. (b)
     $P(t)$ for a muon tetrahedrally coordinated to four $I=7/2$ nuclei, all subject to the EFG
     generated by the muon itself.}
     \label{fig:example}
 \end{figure}

\begin{equation*}
\omega_{D}=\frac{\mu_{0}}{4 \pi} \frac{\gamma_{\mu} \gamma_{I} \hbar}{r_{I}^{3}}, \qquad  \omega_Q = \frac{e V_{zz} Q}{4 I\left(2 I-1\right)\hbar} .
\end{equation*}

Fig.~\ref{fig:example}a shows the muon polarization as a function of time
for various values of $\omega_{Q}/\omega_{D}$ for  a single nuclear spin $I=7/2$. 
This simple model illustrates how, in the two extreme regimes of zero and
large quadrupolar splitting, the classical expectation of a single
precession frequency is recovered, while, in intermediate regimes, multiple frequencies appear.
Similarly, a departure from the semicalssical KT behaviour can also
be appreciated in the more relevant case of a muon generating an EFG on
four tetrahedrally coordinated $I=7/2$ nuclei. 
The polarization as a function of time is obtained numerically in
this case and shown in Fig.~\ref{fig:example}b.
The trend recovers the 1/3 tail of the classical
KT limit only in the small and large quadrupolar splitting conditions,
while substantial deviations happen for the intermediate regime.

While the details of the muon polarizations in A15 compounds are connected to the peculiar composition
of intrinsic and muon induced EFGs at the X sites, yielding to the simulated curves of Fig.~\ref{eq:model},
the behavior can be qualitatively understood considering the ratio $|\omega_{Q}|/\omega_{D}$ for the nn.
Indeed this ratio happens to be about 2.4 for \Nb, 2 for \Sn, and 0.2 for \Si,{}
thus qualitatively explaining the deviations from a KT-like trend of the latter two samples.

In conclusion, we have presented the experimental observation of coherent oscillations originating
from the interaction between the muon and nuclei with $I=7/2$.
This signal is analogous to what has already been observed in fluorides
and other materials containing $I=1/2$ nuclei with high nuclear moments.
An accurate description of the \mSR spectra was obtained by solving
parameter free spin Hamiltonians that consider the perturbed EFG at nuclear
sites surrounding the muon and effectively include all nuclear spins in the
system to correctly describe long-time depolarization.
In \mSR experiments the time evolution of the muon spin polarization depends
dramatically upon the electronic distribution at
quadrupolar nuclei coupled to the muon and an accurate estimation of the
perturbed EFG at these sites is crucial for a successful analysis.
We have shown that DFT based simulations can be effectively used to this
aim and how their combination with simple spin Hamiltonians represents a
computationally inexpensive method to accurately predict the \mSR spectra
of nuclear origin in virtually any crystalline material.
The strong dependence of the \mSR signal on the EFGs and the possibility of
estimating quantitatively the perturbation of an interstitial $\mu^{+}$ opens
the possibility of using positive muons as a quantum sensing tool to
probe also charge related phenomena in materials.

\acknowledgments
The authors acknowledge Pascal Lejay for providing \Si and \Sn samples.
This work is based on experiments performed at the Swiss Muon Source SµS, Paul Scherrer Institute, Villigen, Switzerland.
We gratefully acknowledge the Science and Technology Facilities Council (STFC) for access to muon beamtime at the
ISIS Neutron and Muon Source (EMU facility).  S. S. thanks Consiglio Nazionale delle Ricerche (CNR) for supporting his visit at ISIS.
We thank René Flükiger for support for sample selection, Franz Lang and Peter Baker for support during the experiment at ISIS.
V. F. M. acknowledges support of the National Science Foundation grant \#DMR-1905532.
P. B., M. M. I and R. D. R. acknowledge funding from the SUPER (Supercomputing Unified Platform - Emilia-Romagna) regional project and computing power form STFC's SCARF cluster.
\bibliography{main}

\end{document}